\documentclass[11pt,letterpaper]{article}
\usepackage{float}
\usepackage{graphicx}
\usepackage{braket}
\usepackage{amsmath}
\usepackage{amsthm}
\usepackage{amsfonts}
\usepackage[english]{babel}
\newtheorem{theorem}{Theorem}

\newtheorem{definition}{Definition}
\newtheorem{proposition}{Proposition}

\newtheorem{remark}{Remark}
\newtheorem{example}{Example}
\usepackage{algorithm}
\usepackage{algorithmic}

\usepackage{hyperref}
\usepackage[margin=1.0in]{geometry}
\usepackage{setspace}
\usepackage{authblk}

\newcommand{\DOI}[1]{doi: \href{http://engine.scichina.com/doi/#1}{#1}}
\begin{document}
	
	\begin{spacing}{1.0}
		\bibliographystyle{plain}
	\title{Quantum Speedup of Twin Support Vector Machines}
	\author{Zekun Ye$^{1, 3}$}
	\author{Lvzhou Li$^{1,4,}$\thanks{lilvzh@mail.sysu.edu.cn\\ \DOI{10.1007/s11432-019-2783-7} - Published 27 February 2020} }
	\author{Haozhen Situ$^{2, 3}$}
	\author{Yuyi Wang$^{1}$}
	\affil{%
		$^1$ School of Data and Computer Science, Sun Yat-Sen University, Guangzhou {\rm 510006}, China
	}%
	
	\affil{%
		$^2$ College of Mathematics and Informatics, South China Agricultural University, Guangzhou {\rm 510642}, China
	}%
	\affil{%
		$^3$ Center for Quantum Computing, Peng Cheng Laboratory, Shenzhen {\rm 518055}, China
	}%
	\affil{%
		$^4$ Ministry of Education Key Laboratory of Machine Intelligence and Advanced Computing (Sun Yat-sen University), Guangzhou  {\rm 510006}, China
	}%
	\maketitle
		Machine learning enables agents or computers to learn from data and improves the performance of specific tasks without requiring explicit programming. As the amount of available data continues to grow, existing machine learning algorithms cannot satisfy the efficiency requirements of various practical applications, and several methods have been proposed to solve this problem, e.g., distributed computing. However, distributed computing, even in the best case, cannot achieve exponential acceleration. Additionally, there are typically extra communication costs. 
		Quantum computers are not yet available for practical applications; however,  efficient quantum algorithms have been proposed to address machine learning problems. 
		
		In this paper, we provide further evidence that quantum computing has the potential to increase the speed of machine learning algorithms in the future.
		The basic unit of quantum information, i.e., qubits, could be in superposition states; quantum computers are expected to be very advantageous for processing high-dimensional data. Thus, the performance of certain machine learning models and algorithms could be significantly improved using quantum computers in terms of time and space complexity.
		
		We devise new quantum algorithms that exponentially speeds up the training and prediction procedures of twin support vector machines (TSVM). To train TSVMs using quantum methods, we demonstrate how to prepare the desired input states according to classical data, and these states are used in the quantum algorithm for the system of linear equations. In the prediction process, we employ a quantum circuit to estimate the distances from a new sample to the hyperplanes and then make a decision. The proposed quantum algorithms  can learn two non-parallel hyperplanes and classify a new sample by comparing the distances from the sample to the two hyperplanes in $O(\log mn)$ time, where $m$ is the sample size and $n$ is the dimension of each data point. In contrast, the corresponding classical algorithm requires polynomial time for both the training and prediction procedures.

		\section{TSVMs model}
		\label{TSVM}
		In the binary classification problems, there are $m$ training examples sampled from some fixed but unknown distribution, including $m_1$ positive examples and $m_2$ negative examples, where each training example is a real vector. Herein, we use a matrix $A \in \mathbb{R}^{m_1 \times n}$ to represent examples of the positive class, and each matrix row represents a positive example. Similarly, a matrix $B \in \mathbb{R}^{m_2 \times n}$ represents examples of the negative class. When a new example, $\vec{x}$, arrives from the same distribution, we can classify it using information extracted from the training examples. TSVMs are widely used binary classification models that attempt to find two non-parallel hyperplanes:
		\begin{equation}
		\vec{w_{1}} \cdot \vec{x} + b_1 = 0, \label{hypereq1}
		\ \vec{w_{2}} \cdot \vec{x} + b_2 = 0,
		\end{equation}
		where $w_1,w_2 \in \mathbb{R}^n$ and $b_1,b_2 \in \mathbb{R}$, such that the positive samples are as close as possible to the first hyperplane and distant from the second. Negative samples are as close as possible to the second hyperplane and distant from the first \cite{Jayadeva2007twin}.
		Formally, TSVMs solve two quadratic programming problems with an objective function corresponding to one class and constraints corresponding to the other class.
		Compared with traditional SVMs, TSVMs are faster and have better generalizability.
		Additionally, to simplify quadratic programming problems, inequality constraints are replaced by equality constraints in least square TSVMs \cite{kumar2009least}:
		\begin{align}
		\min_{\vec{w_1},b_1} &\quad \frac{1}{2} {||A \vec{w_1} + b_1\vec{e_1} ||}^2+\frac{1}{2}c_1 {|| \vec{\xi_2}||}^2,\label{o1} \\
		s.t. &\quad -(B \vec{w_1}+b_1\vec{e_2})+\vec{\xi_2} = \vec{e_2}, \label{c1}\\
		\min_{\vec{w_2},b_2} &\quad \frac{1}{2} {||B \vec{w_2} + b_2\vec{e_2} ||}^2+\frac{1}{2}c_2 {|| \vec{\xi_1}||}^2,\label{o2}\\
		s.t. &\quad (A \vec{w_2}+b_2\vec{e_1})+\vec{\xi_1} = \vec{e_1}, \label{c2}
		\end{align}
		where $c_1,c_2$ are penalty parameters, $\vec{e_1},\vec{e_2} $ are all-one column vectors, and $\vec{\xi_1}, \vec{\xi_2}$ are non-negative slack variables. Here, let $L = \min_{\vec{w_1},b_1} \frac{1}{2} {||A \vec{w_1} + b_1\vec{e_1} ||}^2+\frac{1}{2}c_1 {|| \vec{\xi_2}||}^2$. By substituting the equality constraint \eqref{c1} into the objective function \eqref{o1}, we obtain $L= \min_{\vec{w_1},b_1} \quad \frac{1}{2} {||A \vec{w_1} + b_1\vec{e_1} ||}^2+\frac{1}{2}c_1 {|| B \vec{w_1}+b_1\vec{e_2}+\vec{e_2}||}^2$. In addition, by setting the partial derivatives of the function to zero, we obtain the following:
		\begin{eqnarray*}
			A^T(A \vec{w_1}+b_1\vec{e_1})+c_1 B^T(B\vec{w_1}+b_1\vec{e_2}+\vec{e_2})=0,\label{eq1}\\
			{\vec{e_1}}^T(A \vec{w_1}+b_1\vec{e_1})+c_1 {\vec{e_2}}^T(B\vec{w_1}+b_1\vec{e_2}+\vec{e_2})=0,\label{eq2}
		\end{eqnarray*}
		which can be rewritten as follows:
		\begin{equation}
		\begin{pmatrix}
		\vec{w_1} \\
		b_1
		\end{pmatrix}
		= -(\frac{1}{c_1}E^T E+F^TF)^{-1}F^T\vec{e_2}, \label{eq3}
		\end{equation}
		where $E=[A \ \vec{e_1}],F=[B \ \vec{e_2}]$. Here, 
		Similarly, we obtain the following:
		\begin{equation}
		\begin{pmatrix}
		\vec{w_2} \\	b_2
		\end{pmatrix}	= (E^TE+\frac{1}{c_2}F^TF)^{-1}E^T\vec{e_1}. \label{eq4}
		\end{equation}
		
		Matrix multiplication requires at least $\Omega(n^2)$ time to calculate $F^T F$ and $E^T E$, find the inverse of $\frac{1}{c_1}E^TE+F^TF$ and $E^TE+\frac{1}{c_2}F^TF$, and
		calculate $F^T \vec{e_2}$ and $E^T \vec{e_1}$. Thus, the total running time required to solve the two hyperplanes in Eq. \eqref{o1}-\eqref{c2} is $\Omega(n^2)$. To classify a new instance, we must calculate the distance from the new instance $\vec{x}$ to the two hyperplanes $|\vec{w_1}\vec{x}+b_1|/||\vec{w_1}||$ and $|\vec{w_2}\vec{x}+b_2|/||\vec{w_2}||$, and then compare them. Note that this step requires $\Omega(n)$ time.

		\section{Quantum Algorithms for TSVMs}
		\label{QTSVM}
		Herein, we present the proposed quantum algorithms to increase the speed of the training and prediction procedures of TSVMs (Appendix A provides some background knowledge of quantum computation).
		Without the loss of generality, we assume $||\begin{pmatrix}
		\vec{w_i} \\ b_i
		\end{pmatrix}|| = 1$ for $i=1, 2$, 
		where $||\cdot||$ denotes the $2$-norm of vectors. 
		We introduce the following notations:
		\begin{eqnarray*}
			\ket{F^T\vec{e_2}} =& \frac{F^T\vec{e_2}}{{||F^T\vec{e_2}||}}, \ket{E^T\vec{e_1}} = \frac{E^T\vec{e_1}}{{||E^T\vec{e_1}||}}, \\
			\ket{\vec{w_1},b_1} =& \begin{pmatrix}
				\vec{w_1} \\ b_1
			\end{pmatrix}, \ket{\vec{w_2},b_2} = \begin{pmatrix}
				\vec{w_2} \\ b_2
			\end{pmatrix}.
		\end{eqnarray*}
		For $\vec{x} = (x_0,..., x_{n-1})$, we define $\tilde{x} = (x_0,..., x_{n-1}, 1)$ and $\ket{\tilde{x}}=\frac{1}{\sqrt{N_{\tilde{x}}}}\left(\sum_{i=0}^{n-1}x_i\ket{i}+1\ket{n}\right)$, where $N_{\tilde{x}}=\sum_{i=0}^{n-1}x_i^2+1$. 
		Then, we rewrite Eq.\ \eqref{eq3} and Eq.\eqref{eq4} in the quantum setting as follows:
		\begin{equation}
		\ket{\vec{w_1},b_1}
		= \frac{(\frac{1}{c_1}E^TE+F^TF)^{-1}\ket{F^T\vec{e_2}}}{||(\frac{1}{c_1}E^TE+F^TF)^{-1}\ket{F^T\vec{e_2}}||}, \label{eq5} \\
		\end{equation}
		\begin{equation}
		\ket{\vec{w_2},b_2}
		=
		\frac{(E^TE+\frac{1}{c_2}F^TF)^{-1}\ket{E^T\vec{e_1}}}{||(E^TE+\frac{1}{c_2}F^TF)^{-1}\ket{E^T\vec{e_1}}||}. \label{eq6}
		\end{equation}
		In the training procedure, we prepare quantum states $\ket{F^T\vec{e_2}}$ and $\ket{E^T\vec{e_1}}$, and then solve Eq.\ \eqref{eq5} and Eq. \eqref{eq6} by calling the quantum linear system algorithm \cite{Harrow2009quantum} and saving the solutions in quantum states $\ket{\vec{w_1},b_1}$ and $\ket{\vec{w_2},b_2}$ that represent two hyperplanes. In the classification procedures, we represent a new sample $\vec{x}$ as a quantum state $\ket{\vec{x}, 1}$ and obtain the distances between the sample and hyperplanes by estimating the inner product of $\ket{\vec{x},1}$ with $\ket{\vec{w_1},b_1}$ and $\ket{\vec{w_2},b_2}$, respectively.
		

		
		The algorithmic procedures are shown in Algorithm \ref{learning algorithm} and \ref{predicting algorithm}, which are explained in detail as follows. 
		\begin{algorithm}[H]
			\caption{QTSVM training process.}
			\label{learning algorithm}
			\textbf{Input:} $m_1$ positive samples and $m_2$ negative samples represented by matrices $A$ and $B$, where $A \in \mathbb{R}^{m_1 \times n}$ and $B \in \mathbb{R}^{m_2 \times n}$.
			
			\textbf{Procedure:}
			\begin{enumerate}
				\item Prepare input quantum states $\ket{F^T\vec{e_2}}$  and $\ket{E^T\vec{e_1}}$, where $E=[A \ \vec{e_1}]$ and $F=[B \ \vec{e_2}]$. \label{procedure1}
				\item Use the quantum algorithm for systems of linear equations as a subroutine to solve the linear equations shown in Eq. (\ref{eq5}) and (\ref{eq6}), and obtain quantum states $\ket{\vec{w_1},b_1}$  and $\ket{\vec{w_2},b_2}$ that represent two hyperplanes. \label{procedure2}
			\end{enumerate}
			\textbf{Output:} Quantum states $\ket{\vec{w_1},b_1}$  and $\ket{\vec{w_2},b_2}$ that represent two hyperplanes.
		\end{algorithm}
		\begin{algorithm}[H]
			\caption{QTSVM predicting process.}
			\label{predicting algorithm}
			\textbf{Input:} Quantum states $\ket{\vec{w_1},b_1}$ and $\ket{\vec{w_2},b_2}$ that represent two hyperplanes. A new sample $\vec{x} \in \mathbb{R}^n$.
			\textbf{Procedure:}
			\begin{enumerate}
				\item Prepare new sample $\vec{x}$ as a quantum state $\ket{\vec{x},1}$.
				\item Use the SWAP  test to find the distances from new sample to two hyperplanes respectively and compare them. \item If $\vec{x}$ is closer to the first hyperplane, then label it  positive; otherwise negative.
			\end{enumerate}
			\textbf{Output:} The label of $\vec{x}$.
		\end{algorithm}
		Additionally, Theorem 1 gives the time complexity of the proposed algorithms (Appendix D provides the proof of Theorem 1).
		\begin{theorem}
			The time complexities of Algorithm \ref{learning algorithm} and  \ref{predicting algorithm} are $O(\log mn)$ and $O(\log n)$, respectively, where $m=m_1+m_2$.
		\end{theorem} 
		Therefore, for a new sample, we first use Algorithm 1 to learn the  two hyperplanes given by Eq. (\ref{hypereq1}) and then use Algorithm 2 to classify it. The total time is $O(\log mn)$.
		
		\section{Training process}
		First, we provide some notations for Algorithm \ref{learning algorithm}. Let $K_1 = E^T E, K_2 = F^T F, H_1=\frac{1}{c_1}K_1+K_2, H_2=K_1+\frac{1}{c_2}K_2$, and $\hat{K_1} = \frac{K_1}{tr(K_1)},\hat{K_2} = \frac{K_2}{tr(K_2)},\hat{H_1} = \frac{H_1}{tr(H_1)},\hat{H_2} = \frac{H_2}{tr(H_2)}$. Note that Appendix B provides the method to prepare quantum states $\ket{F^	T\vec{e_2}}$ and $\ket{E^T \vec{e_1}}$ based on quantum oracles \cite{ Rebentrost2014quantum}. In the quantum setting, it assumes that oracles for the training data return corresponding quantum states. 
		One way to efficiently construct these states is via quantum RAM that uses $O(mn)$ hardware resources but only $O(\log mn)$ operations to access them \cite{Giovannetti2008quantum}.
		
		To apply the quantum algorithm for systems of linear equations to solve Eq.\ \eqref{eq3} and Eq. \eqref{eq4}, $H_1$ and $H_2$ must be exponentiated efficiently. In other words, it is an important subroutine to calculate $e^{-i\hat{H_1}\Delta t}$ and $e^{-i\hat{H_2}\Delta t}$ efficiently, where $\hat{H_1}$ and $\hat{H_2}$ are the normalization of $H_1$ and $H_2$, respectively,  and $\Delta t$ is a small time slice. Unlike the original method in the quantum algorithm for systems of linear equations, we employ a density matrix exponentiation method \cite{Lloyd2014quantum} herein  that allows us to perform Hamiltonian simulation effectively even if the samples do not satisfy the sparsity assumption. Details are given in Appendix C. 
		We then obtain the two hyperplanes in the form of quantum states $\ket{\vec{w_1},b_1}$ and $\ket{\vec{w_2},b_2}$ by calling the quantum algorithm for the systems of linear equations \cite{Harrow2009quantum}. 
		
		\section{Prediction process}
		In Algorithm \ref{predicting algorithm}, given a new sample $\vec{x}\in \mathbb{R}^n$, we determine its label by comparing the distances from it to the two obtained hyperplanes. By calling the oracle to the vector $\tilde{x}$, we can construct state $\ket{\tilde{x}}$. Using the SWAP test \cite{Buhrman2001quantum}, we then estimate the square of inner product
		$I = {\langle\vec{w_1},b_1|\tilde{x}\rangle}^2=\frac{{|\vec{w_1}\vec{x}+b_1|}^2}{N_{\tilde{x}}}$ by using $O(\epsilon^{-2})$ copies of $\ket{\vec{w_1},b_1}$ and $\ket{\tilde{x}}$ such that the error of the inner product estimation is no greater than $\epsilon$ with high probability.
		
		Additionally, to estimate the value of $||\vec{w_1}||^2$, we employ a project measurement operations set $\{P_i\}$ to measure $\ket{\vec{w_1},b_1}$, where $P_0 = I-\ket{n}\bra{n}, P_1 = \ket{n}\bra{n}$. Then, the probability of obtaining outcome $0$ is $||\vec{w_1}||^2$. By measuring $O(\epsilon^{-2})$ copies of $\ket{\vec{w_1},b_1}$ repeatedly, we can estimate the value of $||\vec{w_1}||^2$ no more than $\epsilon$ with high probability. Similarly, we can estimate the value of $||\vec{w_2}||^2$.
		
		Now, we can obtain the value of ${|\vec{w_1}\vec{x}+b_1|}^2/{{||\vec{w_1}||}^2}$ because the values of ${|\vec{w_1}\vec{x}+b_1|}^2$ and ${||\vec{w_1}||}^2$ have been estimated. Note that the value of ${|\vec{w_2}\vec{x}+b_2|}^2 / {{||\vec{w_2}||}^2}$ can be obtained in a similar manner. If ${|\vec{w_1}\vec{x}+b_1|}^2/{{||\vec{w_1}||}^2} <
		{|\vec{w_2}\vec{x}+b_2|}^2 / {{||\vec{w_2}||}^2} $, then the sample is closer to the first hyperplane, and it will be labeled as a positive point; otherwise, it will be labeled negative.

		\section*{Acknowledgements}{This work is supported by the National Natural Science Foundation of China
			(No. 61772565), the Natural Science Foundation of
			Guangdong Province of China (No. 2017A030313378), the Science
			and Technology Program of Guangzhou City of China (No. 201707010194), and Key R \& D project of Guangdong Province (Grant. 2018B030325001).}

		\clearpage
		\begin{appendix}
			
			\section{Background Knowledge of Quantum Computation}
			We review important notions and notations of quantum computation. Bra-Ket notations $\bra{\cdot}$ and $\ket{\cdot}$ are used to denote vectors. $\bra{v}$ represents the row vector $(v_1,v_2,\ldots, v_n)$,  and $\ket{v}$ the conjugate transpose of $\bra{v}$, i.e., $\ket{v} = \bra{v}^\dagger$.
			
			We now describe the four postulates of quantum mechanics, and a more comprehensive description on quantum computation and quantum information can be found in \cite{Nielsen2000quantum}.	Firstly, a quantum system is associated with a Hibert space and its state is a unit vector in the space. The simplest quantum system is a qubit, the basic unit of quantum information, which lies in a two-dimensional state space. A qubit can be in a superposition state $\ket{\psi} = \alpha\ket{0}+\beta\ket{1},$ where $\alpha, \beta \in \mathbb{C}$, and ${|\alpha|}^2+{|\beta|}^2 = 1$. Moreover, $\ket{0}$ and $\ket{1}$ are called  basis states which correspond to the classical bit $0$ and $1$, respectively, and usually $\ket{0} = \begin{pmatrix} 1 \\ 0 \end{pmatrix},\ket{1} = \begin{pmatrix} 0 \\ 1\end{pmatrix}$. For a $d$-dimensional quantum system, its state space is $\mathbb{C}^d$, and its state is written as $|\psi\rangle=\sum_{i=0}^{d-1}\psi_i|i\rangle,$ where $|i\rangle=(0,...,0,1,0,...,0)^T$ denotes a column vector with the $(i+1)$-th entry being $1$ and else $0$, $\psi_i\in \mathbb{C}$, and $\sum_{i=0}^{d-1}|\psi_i|^2=1$.
			
			Secondly, the evolution of a closed quantum system is described by a unitary operator.  A operator $U$ is said to unitary if $UU^{\dagger}=U^{\dagger} U=I$, where $\dagger$ denotes the conjugate transpose. If the state of the system is $\ket{\psi_1}$ at time $t_1$, and the state of the system is $\ket{\psi_2}$ at time $t_2$, then there exists a unitary operator $U$ which depends only on the time $t_1$ and $t_2$, such that $\ket{\psi_2} = U\ket{\psi_1}.$
			
			Thirdly, we use the quantum measurement to obtain the information from quantum states. An important and common class of measurements is projective measurements.  A projective measurement is described by a set of projective operators  $\{P_i\}$, which satisfy the constraints $\sum_{i}P_i = I$ and $P_iP_j = P_i\delta_{ij}$. If the current state of a quantum system is $\ket{\psi}$, after the measurement, the outcome $m$ will be observed with probability $p(m) = \bra{\psi}P_m \ket{\psi}, $ and the state becomes $P_m\ket{\psi}/{\sqrt{p(m)}}$ after the measurement correspondingly. For example, we measure the quantum state $\ket{\psi} = \alpha\ket{0}+\beta\ket{1}$ by the measurement defined by two projective operators $P_0 = \ket{0}\bra{0}, P_1 = \ket{1}\bra{1}$. Then the probability of obtaining the outcome $0$ is $|\alpha|^2$, and the state becomes $\ket{0}$. Similarly, the probability of obtaining $1$ is $|\beta|^2$, and then the state becomes $\ket{1}$.
			
			Fourthly, the state space of a composite quantum system is the tensor product of the state spaces of the subsystems. If we have $n$ quantum systems, the states of which are $\ket{\psi_1},\ket{\psi_2},...,\ket{\psi_n}$, respectively, then the joint state of the total system is $\ket{\psi_1}\otimes\ket{\psi_2}\otimes...\otimes\ket{\psi_n}$, abbreviated as $\ket{\psi_1}\ket{\psi_2}...\ket{\psi_n}$.
			
			\section{Preparation of Input Quantum States}
			
			We show the preparation of quantum state  $\ket{F^T\vec{e_2}}$ in the following alogrithm. 
			
			\begin{algorithm}[H]
				\caption{Preparation of input quantum states of QTSVM.}
				\label{state preparation}
				\textbf{Input:} Matrix $F \in \mathbb{R}^{m_2*(n+1)}$ with each row $F_i$ stored in quantum RAM by the method we mention above.
				
				\textbf{Procedure:}
				\begin{enumerate}
					
					
					\item Similar to the method in \cite{Rebentrost2014quantum}, call the training data oracles with the state $\frac{1}{\sqrt{m_2}} \sum_{i=0}^{m_2-1}\ket{i}$ to prepare the state 
					$$
					\ket{\chi} = \frac{1}{||F||}\sum_{i=0}^{m_2-1}||F_i||\ket{F_i}\ket{i}.
					$$
					
					
					\item Perform the Walsh-Hadamard transformation to the second register of $\ket{\chi}$ to get state
					$$\frac{1}{||F||}\sum_{i=0}^{m_2-1}||F_i||\ket{F_i}\sum_{j=0}^{m_2-1}(-1)^{ij}\ket{j}.$$
					
					\item Measure the second register. If the measurement result is even (this happens with probability $50\%$), then the state of first register is $\frac{1}{||F||}\sum_{i=0}^{m_2-1}||F_i||\ket{F_i}$. Otherwise, repeat the entire procedure until the measurement result is even.
					
				\end{enumerate}
				\textbf{Output:} Quantum state $ \frac{1}{||F||}\sum_{i=0}^{m_2-1}||F_i||\ket{F_i}$.
			\end{algorithm}
			
			Since $$ 
			\frac{1}{||F||}\sum_{i=0}^{m_2-1}||F_i||\ket{F_i}
			= \frac{1}{||F||}\sum_{i=0}^{m_2-1}||F_i||(\frac{F_i}{||F_i||})^T 
			= \frac{1}{||F||}\sum_{i=0}^{m_2-1}F_i^T,
			$$
			and
			$$
			\ket{F^T\vec{e_2}}
			= \frac{F^T\vec{e_2}}{{||F^T\vec{e_2}||}}
			= \frac{\sum_{i=0}^{m_2-1}F_i^T}{||\sum_{i=0}^{m_2-1}F_i^T||}\\
			= \frac{1}{||F||}\sum_{i=0}^{m_2-1}F_i^T = \frac{1}{||F||}\sum_{i=0}^{m_2-1}||F_i||\ket{F_i},
			$$
			we can prepare $\ket{F^T\vec{e_2}}$ by running Algorithm \ref{state preparation}.
			Similarly, we can prepare $\ket{E^T \vec{e_1}}$.
			
			\section{Density matrix exponentiation}
			We prepare a copy of $\ket{\chi}$ in Algorithm \ref{state preparation} and perform a partial trace operation on the first register to get the density operator
			\begin{align*}
			\mathrm{tr}_1 \{\ket{\chi}\bra{\chi}\} &=\frac{1}{||F||^2}\sum_{i,j=0}^{m_2-1}||F_i||\cdot||F_j||\langle F_j | F_i \rangle \ket{i}\bra{j} \\ &= \frac{F^T F}{\mathrm{tr} (F^T F)} 
			= \hat{K_2},
			\end{align*}
			which needs $O(\log m_2 n)$ time. Similarly, we can prepare $\hat{K_1}$ in $O(\log m_1 n)$ time. Because $m>m_1$ and $m>m_2$, the consumed time is $O(\log mn)$. 
			By Trotter's formula \cite{Nielsen2000quantum}, we have
			\begin{eqnarray*}
				e^{-i\hat{H_1}\Delta t} &=& e^{-\frac{i(\frac{1}{c_1}K_1+K_2)}{\mathrm{tr} H_1}\Delta t}\\
				&=&e^{-\frac{1}{c_1}\frac{iK_1\Delta t}{\mathrm{tr} H_1}}e^{-\frac{iK_2 \Delta t}{\mathrm{tr} H_1}}+O({\Delta t}^2) \\
				&=& e^{-i\hat{K_1}\frac{1}{c_1}\frac{\mathrm{tr} K_1}{\mathrm{tr} H_1}\Delta t}e^{-i\hat{K_2}\frac{\mathrm{tr} K_2}{\mathrm{tr} H_1}\Delta t}+O({\Delta t}^2).
			\end{eqnarray*}
			Since $\frac{\mathrm{tr} K_1}{\mathrm{tr} H_1}$, $\frac{1}{c_2}\frac{\mathrm{tr} K_2}{\mathrm{tr} H_1}$ are constant factors, and $\mathrm{tr} K_1$, $\mathrm{tr} K_2$, $\mathrm{tr} H_1$ can be efficiently  estimated \cite{Rebentrost2014quantum}, $e^{-i\hat{H_1}\Delta t}$ can be simulated in $O(\log mn)$ time with $O({\Delta t}^2)$ error. Moreover, $e^{-i\hat{H_2}\Delta t}$ can then be simulated in the same way.
			
			\section{Proof of Theorem 1}
			
			\begin{proof}
				We consider Algorithm 1 first. In Step 1, it needs $O(\log mn)$ time to prepare $\ket{\chi}$ and $O(\log n)$ time to perform the Walsh-Hadamard transformation.
				The probability that the result of measuring the second register is even is $1/2$ in Algorithm \ref{state preparation}, so the expected number of repetitions required for the entire procedure is constant. 
				Thus, the state preparation time is $O(\log mn)$. 
				In Step 2, we call the quantum algorithm for systems of linear equations \cite{Harrow2009quantum} to solve the Eq.\ 8 and Eq.\ 9, and next we analyze the time complexity and error in this procedure. The errors come from Hamitonian simulation and phase estimation. We denote the error in Hamitonian simulation by $\epsilon_h$, the error in phase estimation $\epsilon_p$. In Hamitonian simulation, we denote the total evolution time by $t_0$, and the number of evolution steps $T$. Then the time slice $\Delta t$ of every step satisfies that $\Delta t = \frac{t_0}{T}$. We need to simulate $e^{-i\tau\hat{H_1}\Delta t}$ and $e^{-i\tau\hat{H_2}\Delta t}$ in this algorithm, where $\tau = 0,1,2,...,T-1$. Since operators $e^{-i\hat{H_1}\Delta t}$ and $e^{-i\tau\hat{H_2}\Delta t}$ can be simulated with $O({\Delta t}^2)$ error, operators $e^{-i\tau\hat{H_1}\Delta t}$ and $e^{-i\tau\hat{H_2}\Delta t}$ can be simulated with $O(T{\Delta t}^2)$ error due to the linear accumulation of error. Since $\Delta t = \frac{t_0}{T}$, we have $\epsilon_h = O(T{\Delta t}^2) = O(\frac{t_0^2}{T})$, thus evolution steps must satisfy that  $T=O(\frac{t_0^2}{\epsilon_h})$. Because it needs $O(\log mn)$ time to simulate $e^{-i\hat{H_1}\Delta t}$, the total time of Hamiltonian simulation is $O(T \cdot \log mn) = O(\frac{t_0^2 \log mn}{\epsilon_h})$. Let $\kappa = \max\{\kappa_1, \kappa_2\}$, where $\kappa_1$ and $\kappa_2$ are the condition number of $\hat{H_1}$ and $\hat{H_2}$ respectively. In order to make the error of phase estimation no more than $\epsilon_p$, it needs to satisfy that $t_0=O(\kappa/\epsilon_p)$ \cite{Rebentrost2014quantum}. Then we have $O(\frac{t_0^2 \log mn}{\epsilon_h}) = O(\frac{\kappa^2 \log mn}{\epsilon_h\epsilon^2})$. Finally, we need to repeat the algorithm $O(\kappa)$ times in order to get a constant success probability, so the time of solving the equations in Step 2 is $O(\frac{\kappa^3 \log mn}{\epsilon_h \epsilon_p^2})$. Therefore, the time complexity of Algorithm 1 is $O(\frac{\kappa^3 \log mn}{\epsilon_h \epsilon_p^2})$.
				
				Next, we turn our attention to Algorithm 2. In Algorithm 2, it needs $O(\epsilon^{-2})$ copies of  $\ket{\vec{w_i},b_i}$ and $\ket{\tilde{x}}$ to estimate the values of ${|\vec{w_i}\vec{x}+b_i|}^2$ and ${||\vec{w_i}||}^2$ for $i=1,2$. Since it needs $O(\log n)$ time to construct state $\ket{\tilde{x}}$, the time complexity of Algorithm 2 is $O(\frac{\log n}{\epsilon^2})$. 
			\end{proof}
		\end{appendix}
	\end{spacing}
	\end{document}